\documentclass[twocolumn,showkeys,aps,prd]{revtex4}

\usepackage{dcolumn}
\usepackage{graphicx}
\usepackage{amsmath}
\usepackage{amssymb}
\usepackage{xspace}
\usepackage[linktocpage={true}]{hyperref}

\begin{document}


\title{Feldman-Cousins Confidence Levels -- Toy MC Method}
\author{Till Moritz Karbach}
\affiliation{TU Dortmund, Germany}
\date{\today}



\begin{abstract}

In particle physics, the likelihood ratio ordering principle is
frequently used to determine confidence regions. This method has
statistical properties that are superior to that of other confidence
regions. But it often requires intensive computations involving
thousands of toy Monte Carlo datasets. The original paper by Feldman and
Cousins contains a recipe to perform the toy MC computation. In this
note, we explain their recipe in a more algorithmic way, show its
connection to $1-\rm CL$ plots, and apply it to simple Gaussian
situations with boundaries.

\end{abstract}


\maketitle


\section{Likelihood ratio ordering}

Feldman and Cousins have developed a method~\cite{Feldman:1997qc} to
compute confidence intervals that has well-defined frequentist coverage.
Their method has certain advantages over the classical approach to
compute confidence levels: it avoids the so-called ``flip-flopping''
because it offers a natural transition between upper (or lower) limits
and two-sided confidence intervals. It usually gives a non-empty interval
even in cases, in which an observable is measured to be far in a
forbidden region--a situation in which classical constructions may lead
to empty intervals, that don't contain the true value with absolute
certainty. And the method effectively decouples the goodness-of-fit
confidence for that for an estimated parameter (cf. the instructive
example in Ref.~\cite{yabsley}).

In general, frequentist confidence intervals $[\mu_1, \mu_2]$ of
confidence level (CL) $\alpha$ are computed by constructing the
confidence belt (Neyman construction, cf. Ref.~\cite{pdg}). This belt
consists of the conjunction of the intervals $[x_1(\mu), x_2(\mu)]$,
that is for each value of $\mu$ one finds an interval by integrating the
probability density function $P(x|\mu)$ such that
\begin{equation}
    \label{eq:defcl}
    \int_{x_1}^{x_2} P(x|\mu) \, {\rm d}x = \alpha ~.
\end{equation}
The belt has the property, that as long as each of the intervals
$[x_1, x_2]$ fulfill Eq.~\ref{eq:defcl}, each vertical (that is
in $\mu$ direction) intersection gives an interval of coverage
$\alpha$. Thus, after the belt is constructed, one
determines the confidence interval $[\mu_1, \mu_2]$ by intersecting the
belt at the measured value of $x$.

But Eq.~\ref{eq:defcl} doesn't determine $x_1$ and $x_2$ uniquely.
In order to solve uniquely for $x_1$ and $x_2$ one also needs to specify
an ordering principle that determines in which order the ${\rm d}x$
shall be included into the interval. The classical intervals are
obtained by including the ${\rm d}x$ in order of their probability.
This yields the additional requirement that
\begin{equation}
    P(x_1|\mu) = P(x_2|\mu) ~.
\end{equation}

Feldman and Cousins define an alternative ordering principle. They
include the ${\rm d}x$ in oder of decreasing likelihood ratio,
\begin{equation}
	\label{eq:defr}
    R(x,\mu) = \frac{P(x|\mu)}{P(x|\mu_{\rm best})} ~.
\end{equation}
Here, $\mu_{\rm best}$ is the value of $\mu$ that maximizes the
likelihood $P(x|\mu)$ and at the same time is inside the allowed region.
Note the distinction between probability and likelihood. Although
numerically identical, one refers to $P(x|\mu)$ as probability density
only if interpreted as a function of $x$. If interpreted as a function
of the true parameter $\mu$, it is called likelihood, because in the
frequentist interpretation the true parameter is what it is and it
doesn't make sense to assign a probability to it. In practice, using the
likelihood ratio ordering means that for a given value of $\mu$, one
finds the interval $[x_1, x_2]$ such that
\begin{equation}
    \label{eq:fcordering}
    R(x_1)=R(x_2)
\end{equation}
and Eq.~\ref{eq:defcl} hold. It is worth noting that in the case of
discrete probability densities such as the Poisson distribution, the
equalities of Eqns.~\ref{eq:defcl}-\ref{eq:fcordering} in general 
can't be met exactly.


\section{Toy MC Method by F.-C.}

In their original publication~\cite{Feldman:1997qc} Feldman and Cousins
offer a method to solve Eqns.~\ref{eq:defcl} and~\ref{eq:fcordering}
numerically. The method is based on toy Monte Carlo (MC) experiments,
which are datasets drawn from the assumed p.d.f. They apply the method
to a non-trivial example from neutrino physics. Here we shall use it to
compute the confidence belt for the trivial example of a unit Gaussian
bound to positive values, $\mu>0$.

In practice it is convenient to take the logarithm of the likelihood ratio
\begin{align}
	-2\ln R \equiv \Delta\chi^2 &= \chi^2 - \chi^2_{\rm best} \\
	\label{eq:defdeltachisq}
	&= \chi^2(x,\mu)  - \chi^2(x,\mu_{\rm best}) ~,
\end{align}
because one often prefers to minimize $-2\ln P$ rather than to maximize
$P$. Due to the monotonic nature of the logarithm, neither the position
of the extrema nor the ordering principle change. 

The algorithm proposed by F.-C. computes the interval $[x_1, x_2]$ for
each value  of the true parameter (the mean $\mu$ of the Gaussian) in a
sufficiently fine grid. The confidence belt is the conjunction of all
such intervals. The algorithm is described in turn:

\begin{enumerate}

\item At the considered value of the true parameter $\mu_0$, generate a
toy experiment by drawing a value $x_{\rm toy}$ from the p.d.f. That is,
draw from the unit Gaussian $G(\sigma=1, \mu=\mu_0)$.

\item Compute $\Delta\chi^2_{\rm toy}$ for the toy experiment, following
Eq.~\ref{eq:defdeltachisq}. For $\chi^2(x,\mu)$, $x$ is set  to $x_{\rm
toy}$ and $\mu$ is set to $\mu_0$. For $\chi^2(x,\mu_{\rm best})$,
$x=x_{\rm toy}$ and $\mu_{\rm best} = \max(0, x_{\rm toy})$ implementing
the boundary at zero.

\item Find the value $\Delta\chi^2_c$, such that $\alpha$ of the toy
experiments have $\Delta\chi^2_{\rm toy}<\Delta\chi^2_c$.

\item The interval $[x_1, x_2]$ is given by all values of $x$
such that $\Delta\chi^2(x,\mu_0)<\Delta\chi^2_c$.

\end{enumerate}

Note that the general case quickly gets more expensive than the sketched
example. If there are nuisance parameters present, the likelihood needs
to be minimized in their respect when calculating $\chi^2$ and
$\chi^2_{\rm best}$. Also, the fact that in step 2 we can set $\mu_{\rm
best} = x_{\rm toy}$ if away from the boundary is a consequence of the
$x\leftrightarrow\mu$ symmetry of a Gaussian. In general, $\mu_{\rm
best}$ needs to be obtained from a minimization, that respects the
boundaries of the problem.

\begin{figure}[htb]
\center
\includegraphics[width=0.45\textwidth]{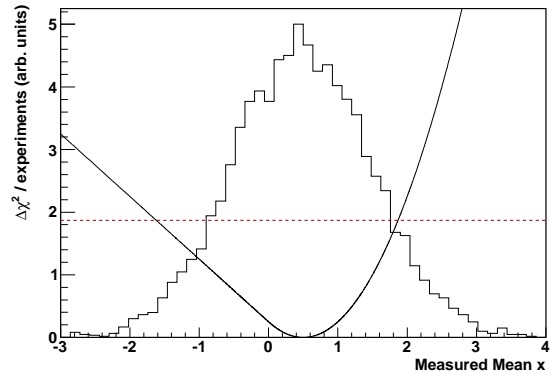}
\caption{Example set of 5000 toy experiments drawn from $G(\mu=0.5,\sigma=1)$
(histogram), together with the likelihood ratio following 
Eq.~\ref{eq:defdeltachisq} computed in presence of a boundary
$\mu>0$ (solid line), and the value $\Delta\chi^2_c$
giving $\alpha=90\%$ (dashed line).}
\label{fig:rc}
\end{figure}

\begin{figure}[htb]
\center
\includegraphics[width=0.45\textwidth]{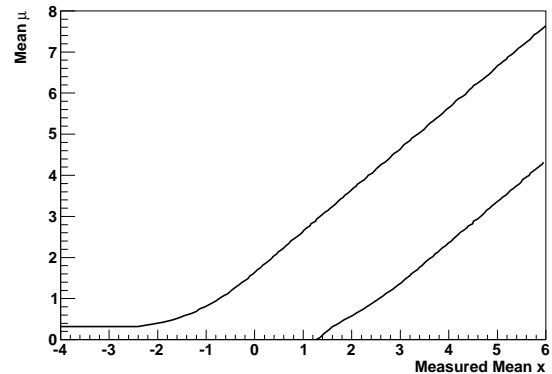}
\caption{Confidence belt for a unit Gaussian with a mean
bound to be positive, $\alpha=90\%$.}
\label{fig:belt}
\end{figure}

The algorithm is illustrated in Fig.~\ref{fig:rc}. It displays all toy
experiments generated in step 1, and the curve built by their
$\Delta\chi^2_{\rm toy}$ values. At values for the measured $x$ greater
than zero, it follows a parabola, $\Delta\chi^2 = (x-\mu)^2$. At values
below zero the effect of the boundary becomes visible when the curve
becomes linear, $\Delta\chi^2 = (x-\mu)^2 - x^2 = \mu^2-2\mu x$. In step
3, an interval is selected such that a fraction $\alpha$ of the toy
experiments fall inside. This satisfies Eq.~\ref{eq:defcl}. The value
$\Delta\chi^2_c$ for which this is the case is depicted in
Fig.~\ref{fig:rc} as a horizontal line. Finally in step 4 the interval
boundaries $x_1$ and $x_2$ are determined as the intersection of the
$\Delta\chi^2_{\rm toy}$ curve and the $\Delta\chi^2_c$ line. This
satisfies Eq.~\ref{eq:fcordering}. The full resulting confidence belt is
shown in Fig.~\ref{fig:belt}.

For illustration, we compute confidence belts for three
additional similar situations: 90\% CL without boundaries,
90\% CL with a two-sided boundary $0<\mu<3$, and 68\% CL
with a boundary at $\mu>0$. The previous example of Fig.~\ref{fig:belt}
is repeated for comparison.

\begin{figure}[htb]
\center
\includegraphics[width=0.45\textwidth]{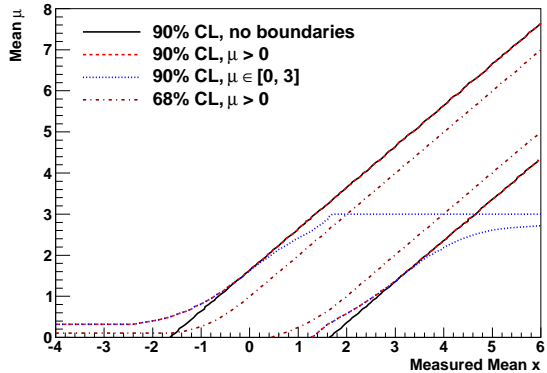}
\caption{Additional example confidence belts for a unit Gaussian.}
\label{fig:beltExamples}
\end{figure}


\section{1--CL plots}
\label{sec:oneminuscl}

\begin{figure}[htb]
\center
\includegraphics[width=0.45\textwidth]{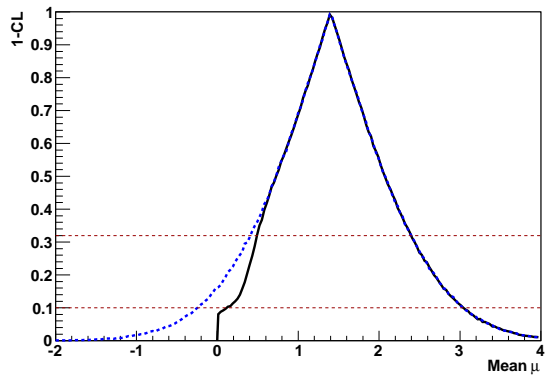}
\caption{Plot of $1-\rm CL$ of a unit Gaussian for a measured mean of
$x=1.4$, both without boundaries (dashed) and for a boundary at $\mu>0$
(solid). The horizontal lines indicate $\alpha=68\%$ and $\alpha=90\%$.}
\label{fig:oneminuscl}
\end{figure}

Confidence belts might not always be the most convenient way to display
information about the statistical precision of a measurement. For
instance, if we performed not only one measurement of the above 
Gaussian mean, the belt becomes a multi-dimensional object. Instead, one
often choses to show ``$1-\rm CL$'' plots for a parameter of interest.
An example is given in Fig.~\ref{fig:oneminuscl}. These plots display
confidence regions for all possible confidence levels, evaluated for one
specific (the observed) measured value $x_0$. To read off the, say, 68\%
CL interval $[\mu_1,\mu_2]$ one draws a horizontal line at $1-\rm CL = 1
- 0.68 = 0.32$. The interval is given by the intersection of this line
and the curve. At the most probable value for the true parameter the
curves raises to one.

The $1-\rm CL$ plots are constructed using a slight variation of the
original toy MC method described above. Steps 1 and 2 remain unchanged:

\begin{enumerate}

\item At the considered value of the true parameter $\mu_0$, generate a
toy experiment.

\item Compute $\Delta\chi^2_{\rm toy} \equiv \Delta\chi^2(x_{\rm toy}, \mu_0)$ for the toy experiment.

\item Compute $\Delta\chi^2_{\rm data} \equiv \Delta\chi^2(x_0, \mu_0)$ for the measured value.

\item The $1-\rm CL$ ($\equiv 1-\alpha(\mu_0)$) value is given by the
fraction of toy experiments that have a larger $\Delta\chi^2$ than the
measured value:
\begin{align}
    \label{eq:oneminuscl}
    1-\alpha(\mu_0) = \frac{N( \Delta\chi^2_{\rm data} < \Delta\chi^2_{\rm toy} )}{N_{\rm toy}} ~.
\end{align}

\end{enumerate}

The full $1-\rm CL$ curve is then obtained by scanning for each possible
value $\mu_0$. The equivalence to the original algorithm can be seen as
follows. The confidence interval of coverage $\alpha_0$ for the true
parameter $\mu$ computed by the $1-\rm CL$ algorithm is given by
\begin{align}
    I_\mu = \left\{ \mu \: \middle| \: 1-\alpha(\mu_0) > 1-\alpha_0 \right\} ~.
\end{align}
In step 3 of the original algorithm we have defined $\Delta\chi^2_c$
such that $\alpha_0$ of the toy experiments have smaller $\Delta\chi^2$
values, thus we can replace both sides of the inequality,
\begin{align}
    \frac{N( \Delta\chi^2(x_0, \mu) < \Delta\chi^2_{\rm toy} )}{N_{\rm toy}} 
    > \frac{N(\Delta\chi^2_c < \Delta\chi^2_{\rm toy} )}{N_{\rm toy}} ~,
\end{align}
leading to
\begin{align}
    I_\mu = \left\{ \mu \: \middle| \: \Delta\chi^2(x_0, \mu) < \Delta\chi^2_c \right\} ~.
\end{align}
Due to the Neyman construction of the confidence belt both horizontal
and vertical intersections of the belt give intervals of coverage $\alpha_0$.
Thus it is also
\begin{align}
    I_x = \left\{ x \: \middle| \: \Delta\chi^2(x, \mu_0) < \Delta\chi^2_c \right\} ~,
\end{align}
which is just the definition of step 4 of the original algorithm.

Two example physics analyses where this method is used can be found
in Refs.~\cite{babarGLW, babarGGSZ}.


\section{Two-dimensional CL contours}

It is straight-forward to apply the toy MC method to a two-dimensional
situation. If there are two, possibly correlated true parameters that
shall be estimated from the data, the $1-\rm CL$ plots become
three-dimensional. From those, CL contours are readily obtained by
slicing at the desired confidence level $1-\alpha_0$.

\begin{figure}[htb]
\center
\includegraphics[width=0.45\textwidth]{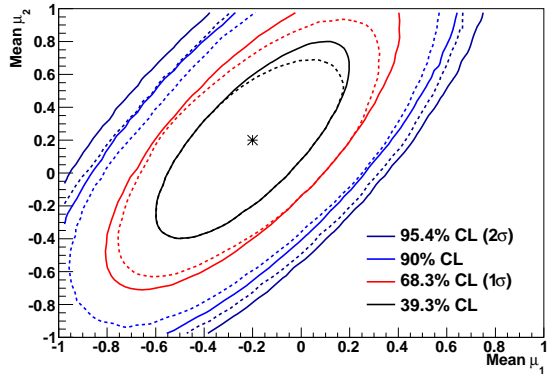}
\caption{Confidence regions for the mean of a two-dimensional Gaussian
with $\sigma_1=0.4$, $\sigma_2=0.6$, $\rho=70\%$, for a measured mean of
$(x_1,x_2)=(-0.2,0.2)$ (data point) with coverage $\alpha=39.3\%$,
68.3\%, 90\%, 95.4\% (solid lines from inner to outer). The dashed lines
show the corresponding regions in presence of the boundaries $|\mu_1|<1$
and $|\mu_2|<1$.}
\label{fig:oneminuscl2d}
\end{figure}

We use the toy MC method to obtain CL contours for the means $\mu_1$ and
$\mu_2$ of a two-dimensional Gaussian. We chose the corresponding widths
$\sigma_{1,2}$ different from each other, and also a non-zero
correlation. The resulting contours are tilted ellipses. They are shown
in Fig.~\ref{fig:oneminuscl2d} (solid lines). The innermost ellipsis has
the property, that its projections onto the $\mu_{1,2}$ axes are exactly
$\pm\sigma_{1,2}$. Its confidence level is $\alpha_0=39.3\%$.

We also consider the effect of boundaries for the true parameters
$\mu_{1,2}$. In Fig.~\ref{fig:oneminuscl2d} (dashed lines), we chose the
allowed region to coincide with the displayed region. The resulting
contours all cover a smaller area compared to the unconstrained case:
The boundaries seem to ``deflect'' the contour, resulting in ``squeezed
ellipses''.

\section{Prob method}

In many cases, the likelihood function assumes a Gaussian shape. This is
likely the case, when the parameter of interest is itself an estimator
that is based on a sufficiently large data sample. Then, $\Delta\chi^2$
follows a $\chi^2$ distribution with one degree of freedom by
definition--which also justifies the use of the symbol in the above. It
is therefore possible to compute $1-\rm CL$ from the cumulative
distribution function of the $\chi^2$ distribution (compare to
Eq.~\ref{eq:oneminuscl}):
\begin{align}
    1-{\rm CL} &= \int_{\Delta\chi^2_{\rm data}}^{\infty} P_{\chi^2}(t) \, {\rm d}t \\
               &= \frac{1}{\sqrt{2} \, \Gamma(1/2)} \int_{\Delta\chi^2_{\rm data}}^{\infty} t^{-1/2} e^{-t/2} \, {\rm d}t \\
               &= \textsc{Prob}(\Delta\chi^2_{\rm data}, 1) ~. \label{eq:prob}
\end{align}
With this approach there is no need for toy MC. The algorithm in
Section~\ref{sec:oneminuscl} just reduces to steps 3 and 4, where
in the latter the toy MC ratio is replaced by Eq.~\ref{eq:prob}.
The \textsc{Prob} function is available within the \textsc{Root}
framework as \texttt{TMath::Prob()}.

It should be stated clearly that this method only applies in the
Gaussian regime. For instance, it does not handle the presence of
boundaries in a satisfactory way. Applied to the situation of
Fig.~\ref{fig:oneminuscl} it produces a curve that coincides with the
unconstrained F.-C. curve everywhere: When reaching the boundary
$\mu=0$, it suddenly drops off to zero. In this situation, correct
coverage is not guaranteed.

The clear advantage of the \textsc{Prob} method is that the required
amount of computation is very reasonable, as no toy MC generation is
involved. And often it gives a very good idea of the F.-C. confidence
contours, so it may be instructive to try it before a F.-C. analysis
is performed.


\section{Conclusion}

In their original publication, Feldman and Cousins include a recipe to
perform a toy MC computation of their confidence regions. In this note, 
we explain it in an algorithmic way, that is suited to direct
implementation into existing fitter frameworks. We show the connection
of the toy MC method to the frequentist confidence belt and to $1-\rm
CL$ plots. We illustrate the method using simple one- and
two-dimensional Gaussian situations with and 
without boundaries on their
true parameters.

\vspace{1cm} 
\section{Acknowledgements}

The author wishes to thank Giovanni Marchiori and Klaus Wacker 
for useful discussion.


\bibliography{Bibliography}{}
\bibliographystyle{h-physrev5}

\end{document}